\documentclass[twocolumn,showpacs,prl]{revtex4}
\usepackage{dcolumn}% Align table columns on decimal point
\usepackage{amsmath,amssymb,epsfig}
\usepackage{paralist}
\usepackage{comment}
\usepackage{graphicx}
\usepackage{multirow}

\begin{document}

\title{Support for the thermal origin of the Pioneer anomaly}
\author{Slava G. Turyshev$^1$, Viktor T. Toth$^2$, Gary Kinsella$^1$, Siu-Chun Lee$^3$, Shing M. Lok$^3$, and Jordan Ellis$^1$}

\affiliation{\vskip 3pt
$^1$Jet Propulsion Laboratory, California Institute of Technology, 4800 Oak Grove Drive, Pasadena, CA 91109-0899, USA
}

\affiliation{\vskip 3pt
$^2$Ottawa, ON  K1N 9H5, Canada
}

\affiliation{\vskip 3pt
$^3$Applied Sciences Laboratory, 13111 Brooks Drive, Suite A,
Baldwin Park, CA 91706-7902, USA
}

\begin{abstract}
We investigate the possibility that the anomalous acceleration of the Pioneer 10 and 11 spacecraft is due to the recoil force associated with an anisotropic emission of thermal radiation off the vehicles. To this end, relying on the project and spacecraft design documentation, we constructed a comprehensive finite-element thermal model of the two spacecraft. Then, we numerically solve thermal conduction and radiation equations using the actual flight telemetry as boundary conditions. We use the results of this model to evaluate the effect of the thermal recoil force on the Pioneer 10 spacecraft at various heliocentric distances. We found that the magnitude, temporal behavior, and direction of the resulting thermal acceleration are all similar to the properties of the observed anomaly. As a novel element of our investigation, we develop a parameterized model for the thermal recoil force and estimate the coefficients of this model independently from navigational Doppler data. We find no statistically significant difference between the two estimates and conclude that once the thermal recoil force is properly accounted for, no anomalous acceleration remains.
\end{abstract}


\maketitle

\section{Introduction}

The anomalous acceleration of the Pioneer 10 and 11 spacecraft \cite{pioprl} is a discrepancy between modeled and observed radio-metric Doppler data received from the two vehicles. The discrepancy can be eliminated by incorporating a constant sunward acceleration of unknown origin with magnitude of $a_P=(8.74\pm1.33)\times10^{-10}~{\rm m/s}^2$ into the orbital model \cite{pioprd}. The presence of this acceleration, which became known as the Pioneer anomaly, was seen by many as a breakdown in the gravitational inverse-square law of general relativity that reveals itself in the dynamics of the outer solar system. There were also proposals that this anomaly was due, at least in part, to the waste heat emitted by the spacecrafts' radioisotope thermoelectric generators (RTGs) \cite{KATZ1998}, their electrical subsystems \cite{MURPHY1998}, or both \cite{LKS2003}, motivating a thorough investigation of the spacecraft systematics (for review, see \cite{piolrr}).

Our current investigation began with the recovery of the entire telemetry record of both spacecraft, substantial project documentation, and additional Doppler data \cite{MDR2005}. The analysis of the extended Doppler data set was completed recently \cite{pio2011a}, confirming the existence of the anomaly. This analysis also showed that although the direction of the anomalous acceleration cannot be determined unambiguously, the Doppler residuals improve if one considers a temporally varying acceleration in the direction of the Earth, which is consistent with earlier, similar results \cite{CBM2005,TOTH2008}. These results are suggestive: If the acceleration were due to thermal recoil force, it would be along the spacecraft spin axis, which generally points in the direction of the Earth and would have temporally decreasing magnitude consistent with the decay of the on-board radioisotope fuel (Pu$^{238}$).

In this {\it Letter} we report on the completion of a finite-element thermal model of the Pioneer 10 spacecraft. This model relies extensively on the  project and spacecraft design documentation and was validated using redundant flight telemetry data. A parameterized form of the thermal force model \cite{Toth2009} was also incorporated into a Doppler analysis, which yielded an independent estimate of coefficients characterizing the thermal recoil force. We compare the outputs of the two independent analyses -- Doppler and thermal -- to show that no statistically significant anomalous acceleration remains in the data.

\vskip -24pt ~

\section{Finite-element thermal model}

\begin{figure}[t]
\includegraphics[width=0.9\linewidth]{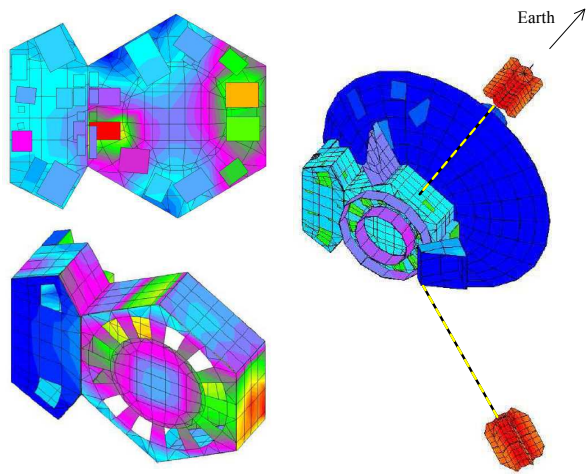}
\caption{\label{fig:view}Illustrative representation of the thermal model of the Pioneer 10 spacecraft evaluated at 40~AU. Top left: spacecraft body interior (temperature range: blue $-16^\circ$~C, red $+10^\circ$~C); Bottom left: spacecraft exterior (blue $-155^\circ$~C, red $-108^\circ$~C); Right: entire spacecraft (blue $-213^\circ$~C, red $+136^\circ$~C). Unmodeled struts that connect the RTGs to the spacecraft body are indicated with yellow-black dashed lines.}
\vskip -12pt
\end{figure}

A comprehensive finite-element (FE) thermal model (Fig.~\ref{fig:view}) of the Pioneer 10 and 11 spacecraft was constructed at the Jet Propulsion Laboratory (JPL) in collaboration with the Applied Sciences Laboratory (ASL). The geometric and thermal models of the spacecraft were constructed using the SINDA/3D thermal modeling software \footnote{Network Analysis Inc., Tempe, Arizona}. While the software provides the capability to build a numerical model directly from CAD drawing files, no such files exist for a spacecraft designed 40 years ago. For this reason, the model was built in a more tedious manner by specifying the coordinates of the vertices of each modeled spacecraft surface, using available blueprints and recovered project documentation. The spacecraft geometric model was built with a Monte Carlo based radiation analyzer (TSS) to calculate the radiative exchange factors using infrared emittance values for modeled surfaces specified within it. The model incorporated approximately 3,300 surface elements, 3,700 nodes, and 8,700 linear conductors. The spacecraft thermal-mechanical configuration is simulated by a network of thermal capacitance, conductive couplings, and radiative exchange factors between all surfaces and to deep space.

\begin{table}[t]
\caption{Pioneer 10 telemetered power at select heliocentric distances. Externally vs. internally located components are indicated where applicable (only 5 out of 11 distances shown).}
\label{tb:power}
\begin{tabular}{l|r|r|r|r|r} \hline\hline
~&\multicolumn{5}{c}{Power (W)}\\\cline{2-6}
Description&~3 AU&10 AU&25 AU&40 AU&70 AU\\\hline\hline
Science, internal&12.6&12.6&11.9&8.8&0.8\\
Science, external&6.4&8.4&6.4&6.4&0.0\\
Subsystems&20.2&20.2&20.4&20.2&19.5\\
Electrical, internal&63.2&46.3&35.5&28.4&17.5\\
Electrical, external&8.1&4.7&2.7&2.3&0.1\\
TWT\footnote{Traveling wave tube amplifier.} thermal&18.6&18.6&19.8&19.5&21.2\\
Transmitter&9.2&9.2&8.0&8.3&6.6\\\hline
\textbf{Total consumed}&\textbf{138.3}&\textbf{120.0}&\textbf{104.7}&\textbf{93.9}&\textbf{65.8}\\\hline\hline
RTG generated&148.5&127.1&107.1&94.0&67.2\\
Cable loss&6.9&5.3&4.0&3.2&1.7\\\hline
\textbf{Total available}&\textbf{141.6}&\textbf{121.8}&\textbf{103.1}&\textbf{90.8}&\textbf{65.5}\\\hline\hline
Difference&$+3.4$&$+1.8$&$-1.6$&$-3.1$&$-0.4$\\
\hline\hline
\end{tabular}
\end{table}

The software numerically solves the energy equation using equipment power dissipation from the spacecraft fight telemetry records (see Table~\ref{tb:power}). RTG power was estimated using the well-known half-life, $\tau=87.74$~yr, of the ${}^{238}$Pu radioisotope fuel: $Q_\mathrm{rtg}(t)=\big[2^{-(t-t_0)/\tau}\big]Q_\mathrm{rtg}(t_0)$, where $t_0=$ July 1, 1972 and $Q_\mathrm{rtg}(t_0)=2578.179$~W. The objective was to calculate the temperature distribution of all spacecraft surfaces. To accommodate the limitations imposed on us by thermal modeling software, the angular distribution of the radiative emission from the spacecraft to space was calculated by solving the thermal radiation equations with the spacecraft positioned at the center of a large (i.e., with radius of 40 high-gain antenna, or HGA, diameters), black spherical control surface. The amount of spacecraft radiative emission absorbed by each element of this control surface corresponds to the amount of momentum carried in this direction.

The model incorporated some parameters the values of which were less well known: e.g., the effective emissivities of multilayer insulation blankets or conductive couplings of certain structural elements. Redundancies in the flight telemetry were used to refine the estimates of these parameters and validate the model. The primary objective was to reproduce the known thermal power of the RTGs by choosing suitable temperature boundary conditions that, in turn, had to agree with telemetered temperature readings. In the final results, the modeled RTG thermal power was within 1\% of the known value, while modeled RTG fin root temperatures were always within $\pm2$~K of the flight telemetry.

Both spacecraft utilized a louver system for thermal management \cite{piolrr}. Louvers mounted on bimetallic springs opened in response to high interior temperatures, allowing excess thermal radiation to escape the spacecraft more freely. Louver geometry for twelve two-blade and two three-blade louver assemblies was integrated into the model. Movements of the modeled blade positions (and the resulting calculated louver effective emittance) were based on the average temperature of the two nodes on the edge of the element that corresponded to the physical location of the actuator housings for each louver assembly.

\begin{figure}[t]
\includegraphics[width=\linewidth]{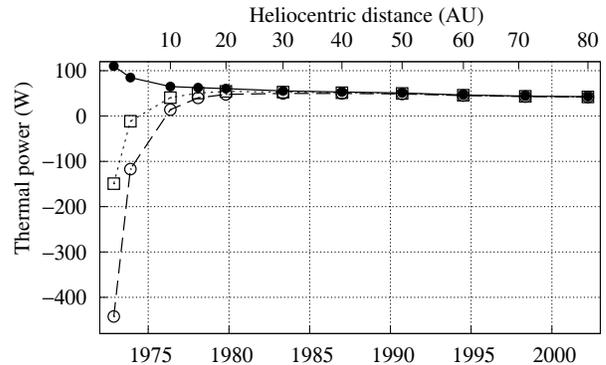}
\vskip -12pt
\caption{Pioneer 10 thermal power contributing to sunward acceleration, including solar heating and reflected solar radiation (hollow circles), solar heating only (hollow squares) with reflected solar radiation removed, and all solar effects removed (filled circles). Positive values indicate radiation directed away from the Sun, resulting in a sunward acceleration of the spacecraft.}
\label{fig:recoil}
\vskip -12pt
\end{figure}

In addition to the internally generated heat, the Sun was also a significant source of heating, particularly at the smaller heliocentric distances.  Since the spacecraft was facing the Sun, the solar energy was absorbed primarily by the HGA which largely shadowed the rest of the spacecraft from direct solar irradiation except for the RTGs. The solar effect became evident as the absorbed energy increased the HGA temperature and was then emitted as IR radiation.

The model was evaluated for Pioneer 10 at eleven different heliocentric distances ranging from 3 AU to 80 AU (see Fig.~\ref{fig:recoil}). Four temperature telemetry locations on the equipment compartment and two on the instrument compartment were examined. The deviation between the predicted values from the model and the flight telemetry was tracked in each of the eleven cases. The differences between telemetered temperature values at the six platform temperature sensors on board vs. values computed from the model reveal a root-mean-square (RMS) error of 5.1\%. This discrepancy is consistent with, and is likely a consequence of, the known sources of error listed in Table~\ref{tb:errorbudget}. Of these, the most significant is the effect of possible surface degradation of the sun-facing RTG surfaces.

\begin{table}[t]
\caption{Error budget for the Pioneer 10 thermal model; the contributions shown in percentages, relative to $a_P$.
}
\label{tb:errorbudget}
\begin{tabular}{r|r} \hline\hline
Description&Error\\\hline\hline
Other sources&$<0.1$\%\\
Quantization error in telemetry data&2.2\%\\
11 1-W radioisotope heater units&2.2\%\\
Discrepancy in TWT telemetry&2.5\%\\
Inaccuracy of geometric model\footnote{Assuming a fore-aft positional uncertainty of 2.5~cm.}&5.0\%\\
Modeling of instrument openings\footnote{Heat escaping through instrument openings is ill-documented.}&5.0\%\\\hline
\textbf{Subtotal:}&\textbf{8.1\%}\\
RTG surface degradation&25.0\%\\\hline
\textbf{TOTAL:}&\textbf{26.3\%}\\ \hline\hline
\end{tabular}
\end{table}

The RTGs were coated with ``three mils of zirconia (ZrO$_2$) in a sodium silicate binder'' \cite{SNAP19}. No information is available in the literature about the performance of this particular type of paint when exposed to solar radiation, especially at the relatively high temperatures present on the RTG outer surfaces. Similar paints \cite{CR1786} have experienced both an increase and a decrease of up to 5\% in infrared emissivity. Approximately 25\% of the RTG coated surfaces were exposed to solar irradiation. A calculation that takes into account the relative contribution of RTG heat to the total anisotropy yields a corresponding error figure of 25\% in the overall error budget.

Other, significant error sources (see Table~\ref{tb:errorbudget}) include the FE modeling uncertainty (obtained by comparing the FE model output to redundant telemetry), insufficient detail in the documentation about the amount of heat escaping through instrument openings, and limited accuracy in the available engineering drawings.

The thermal power $Q$ emitted in a certain direction is related to the acceleration $a_{\rm th}$ resulting from the corresponding thermal recoil force as $a_{\rm th}=\eta \,(Q/mc),$ where $m$ is the mass of the object being accelerated and $c$ is the velocity of light. The dimensionless quantity $\eta$ is an ``efficiency factor''; its value is 1 if the beam is collimated in the direction of the acceleration. It was recognized earlier \cite{Toth2009} that the two main contributions to the thermal recoil force are the waste heat generated by the RTGs, $Q_\mathrm{rtg}$, and the waste heat produced by electrical equipment inside the spacecraft compartments, $Q_\mathrm{elec}$, yielding the anomalous acceleration
\begin{equation}
a_P=\eta_\mathrm{rtg}(Q_\mathrm{rtg}/mc)+\eta_\mathrm{elec}(Q_\mathrm{elec}/mc). \label{eq:Qsum}
\end{equation}
Our earlier investigation \cite{Toth2009} found that the corresponding efficiency factors, $\eta_\mathrm{rtg}$ and $\eta_\mathrm{elec}$, depend only on the geometry of the spacecraft, and remain constant over time; this result was confirmed by the present analysis. The thermal recoil force is well fitted by the model equation using $\eta_\mathrm{rtg}=0.0104$ and $\eta_\mathrm{elec}=0.406$, with a RMS error of only 0.78~W, much smaller than the inherent quantization error present in the telemetry.

\section{Doppler analysis and comparison}

The parameterized thermal recoil force model presented in Eq.~(\ref{eq:Qsum}) can be incorporated easily into the force model used for Doppler analysis. This presented us with the opportunity to go beyond simply using a recoil force estimate in trajectory modeling; the Doppler analysis can also be used to estimate the values of the parameters $\eta_\mathrm{rtg}$ and $\eta_\mathrm{elec}$ completely independently from the thermal analysis.

\begin{table}[t]
\caption{Doppler residuals (in mHz) after incorporating a thermal model, varying the efficiencies of RTG ($\eta_\mathrm{rtg}$) and electrical ($\eta_\mathrm{elec}$) heat conversion into a recoil force.}
\label{tb:doppres}
\begin{ruledtabular}
\begin{tabular}{r|rrrrrrr}
~&\multicolumn{7}{c}{$\eta_\mathrm{rtg}$}\\
$\eta_\mathrm{elec}$&0.0000&0.0096&0.0120&0.0132&0.0144&0.0156&0.0288\\\hline
0.000&~&~&~&~&~&~&5.57\\
0.295&~&10.80&9.03&~&7.37&~\\
0.369&~&8.75&7.11&~&5.70&~\\
0.443&~&6.87&5.50&~&4.62&~\\
0.480&~&~&~&4.57&4.45&4.54\\
0.517&~&~&~&~&4.57&~\\
0.886&4.52&~&~&~&~&~
\end{tabular}
\end{ruledtabular}
\end{table}

Results of this part of our analysis are shown in Table~\ref{tb:doppres}. The spacecraft trajectory was calculated using a variety of values for $\eta_\mathrm{rtg}$ and $\eta_\mathrm{elec}$. The Doppler analysis yielded the lowest residuals at $\eta_\mathrm{rtg}=0.0144$ and $\eta_\mathrm{elec}=0.480$; the residuals, parameterized by $\eta_\mathrm{rtg}$ and $\eta_\mathrm{elec}$, lie on an elongated elliptical paraboloid surface. Consequently, the Doppler analysis is sensitive to the overall magnitude of the recoil force: a $\sim 20\%$ change in the overall magnitude nearly doubles the residual. However, the Doppler analysis cannot disambiguate between the RTG and electrical contributions to the recoil force: an increase in one accompanied by a decrease in the other results in a small or insignificant increase in the Doppler residual, as can be seen from the off-diagonal elements in Table~\ref{tb:doppres}.

\begin{figure}[t]
\includegraphics[width=\linewidth]{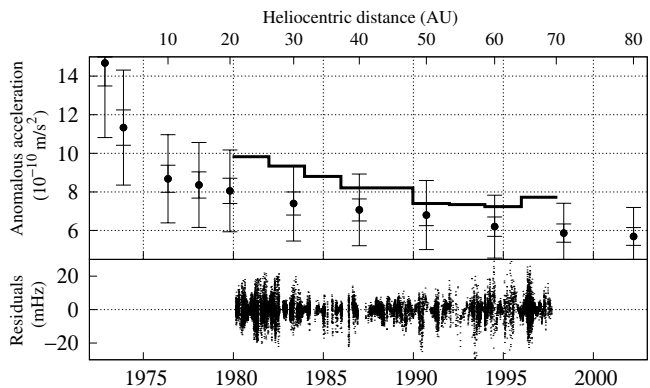}
\vskip -12pt
\caption{\label{fig:result}Comparison of the thermally-induced and anomalous accelerations for Pioneer 10. The estimated thermal acceleration is shown with error bars. The stochastic acceleration estimate from \cite{pio2011a} appears as a step function. For reference, the Doppler residuals of the stochastic acceleration are also shown in the bottom panel. Inner and outer error bars correspond to the subtotal and total shown in Table~\ref{tb:errorbudget}.}
\vskip -12pt
\end{figure}

\begin{figure}[t]
\includegraphics[width=\linewidth]{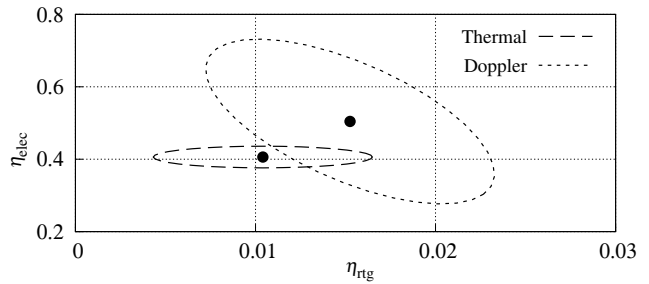}
\vskip -12pt
\caption{\label{fig:errell}Comparison of the RTG ($\eta_\mathrm{rtg}$) and electrical ($\eta_\mathrm{elec}$) efficiency factors estimated independently from the Doppler and thermal analysis using Eq.~(\ref{eq:Qsum}). The 1$\sigma$ error ellipse for the thermal analysis corresponds to the values in Table~\ref{tb:errorbudget}. For the Doppler analysis, the error ellipse corresponds to values of $\eta_\mathrm{rtg}$ and $\eta_\mathrm{elec}$ in Table~\ref{tb:doppres} for which the excess in the Doppler residual exceeds the best-fit residual.
}
\vskip -12pt
\end{figure}

The uncertainty of the thermal recoil force estimate is large. The magnitude of the estimated thermal recoil force is smaller than the observed acceleration, but the stochastic acceleration estimate derived from Doppler data is within 1$\sigma$ (see Fig.~\ref{fig:result}). An error ellipse of the thermal recoil force estimate in the $\eta_\mathrm{rtg}-\eta_\mathrm{elec}$ parameter space of Eq.~(\ref{eq:Qsum}) is shown in Fig.~\ref{fig:errell}. This error ellipse takes into account the fact that the largest source of error (RTG coating) affects only $\eta_\mathrm{rtg}$. Fig.~\ref{fig:errell} also shows the error of the stochastic acceleration, estimated using the best-fit Doppler residual as the 1$\sigma$ noise floor. The orientation of this error ellipse corresponds to Table~\ref{tb:doppres} and reflects the fact that the using the Doppler data alone, the RTG and electrical contributions to the thermal recoil force cannot be well distinguished.

Lastly, we mention that both the thermal recoil force and the Doppler data can be well modeled using an exponential decay model in the form $a=\big[2^{-(t-t_0)/\tau}\big]a_0$. Using $t_0=$ January 1, 1980, the best fit parameters for the Doppler data are \cite{pio2011a} $\tau=(28.8\pm 2.0)$~yr, $a_0=(10.1\pm 1.0)\times 10^{-10}$~m/s$^2$. In contrast, the calculated thermal recoil force can be modeled, with an RMS error of $0.1\times 10^{-10}$~m/s$^2$, using the parameters $\tau=36.9\pm 6.7$~yr, $a_0=(7.4\pm 2.5)\times 10^{-10}$~m/s$^2$.

\section{Conclusions}

We presented results from a thermal analysis of the Pioneer 10 spacecraft and an independently performed analysis of their trajectories using Doppler radio-metric data that also incorporated a parameterized on-board force model.

The comprehensive thermal model yields a recoil force characterized by the parameters $\eta_\mathrm{rtg}=0.0104$ and $\eta_\mathrm{elec}=0.406$. The Doppler analysis yielded the lowest residual at $\eta_\mathrm{rtg}=0.0144$ and $\eta_\mathrm{elec}=0.480$. Numerically, the thermal analysis based estimate of the recoil force is $\sim$80\% (of which $\sim$35\% is produced by the RTGs and $\sim$45\% by electrical heat) of the magnitude estimated from Doppler analysis.

To determine if the remaining 20\% represents a statistically significant acceleration anomaly not accounted for by conventional forces, we analyzed the various error sources that contribute to the uncertainties in the acceleration estimates using radio-metric Doppler and thermal models. On the Doppler side, a significant noise floor exists which is believed to be due to systematics (including the interplanetary charged particle environment, etc.) For the thermal analysis, the biggest source of uncertainty is the unknown change in the properties of the RTG coating, which results in differences in fore-aft emissivity and an additional contribution to the recoil force in the spin axis direction. When we plot these uncertainties in the $\eta_\mathrm{rtg}, \eta_\mathrm{elec}$ parameter space, the 1$\sigma$ error ellipses overlap. We therefore conclude that at the present level of our knowledge of the Pioneer 10 spacecraft and its trajectory, no statistically significant acceleration anomaly exists.

In closing, we must briefly mention additional avenues that may be explored in future studies. First, the case of Pioneer 11 was not analyzed at the same level of detail, albeit we note that spot analysis revealed no surprises for this spacecraft. Second, the question of the anomalous spin-down of both spacecraft remains unaddressed, even though it is plausible that the spin-down is due to heat that is reflected asymmetrically off instrument sunshades. Third, Fig.~\ref{fig:recoil} is strongly suggestive that the previously reported ``onset'' of the Pioneer anomaly  may in fact be a simple result of mismodeling of the solar thermal contribution; this question may be resolved with further analysis of early trajectory data. Fourth, an as-yet unaddressed issue is the possibility of outgassing from surface materials, which was shown to have a potentially observable contribution to the anomalous acceleration \cite{Schlappi2011}. Fifth, our understanding of systematics in the Doppler tracking data can be improved by a detailed autocorrelation analysis. Sixth, the properties of the RTG paint are, in principle, measurable by a thermal vacuum chamber test of a hot RTG analogue. Finally, yet another redundant data set exists in the form of Deep Space Network signal strength measurements, which could be used to improve our understanding of the spacecraft's precise orientation. Nonetheless, it is unlikely that any re-analysis of the Pioneer 10 and 11 data set will alter our main conclusion: the anomalous acceleration of these spacecraft is consistent with known physics.

\section*{Acknowledgments}

We thank G.L. Goltz, K.J. Lee, and N.A. Mottinger of JPL for their indispensable help with the Pioneer Doppler data recovery. We thank W.M. Folkner, T. P. McElrath, M. M. Watkins, and J.G. Williams of JPL for their interest, support, and encouragement. We also thank L.K. Scheffer and C.B. Markwardt for many helpful conversations. We thank The Planetary Society for their long-lasting interest and support. This work in part was performed at the Jet Propulsion Laboratory, California Institute of Technology, under a contract with the National Aeronautics and Space Administration.

\bibliography{new-thermal}

\end{document}